\documentclass[cits]{PoS}

\usepackage{amsmath,amssymb}

\usepackage[square,comma,numbers,sort&compress]{natbib}
\title{Lattice Monte Carlo methods for systems far from equilibrium}

\ShortTitle{Lattice Monte Carlo methods for systems far from equilibrium}

\author{\speaker{David Mesterh\'azy}\\
        Department of Physics, University of Illinois at Chicago\\
        845 West Taylor Street, Chicago IL 60607-7059, USA\\
        E-mail: \email{mesterh@uic.edu}}

\author{Luca Biferale\\
        Dipartimento di Fisica, Universit\`a degli studi di Roma Tor Vergata\\
        Via delle Ricerca Scientifica 1, 00133 Rome, Italy\\
        E-mail: \email{biferale@roma2.infn.it}}

\author{Karl Jansen\\
        NIC/DESY Zeuthen\\
        Platanenallee 6, 15738 Zeuthen, Germany\\
        E-mail: \email{karl.jansen@desy.de}}

\author{Raffaele Tripiccione\\
        Dipertimento di Fisica e Scienze della Terra, Universit\`a degli studi di Ferrara\\
        Via Saragat 1, 44100 Ferrara, Italy\\
        E-mail: \email{tripiccione@fe.infn.it}}

\abstract{We present a new numerical Monte Carlo approach to determine the scaling behavior of lattice field theories far from equilibrium. The presented methods are generally applicable to systems where classical-statistical fluctuations dominate the dynamics. As an example, these methods are applied to the random-force-driven one-dimensional Burgers' equation -- a model for hydrodynamic turbulence. For a self-similar forcing acting on all scales the system is driven to a nonequilibrium steady state characterized by a Kolmogorov energy spectrum. We extract correlation functions of single- and multi-point quantities and determine their scaling spectrum displaying anomalous scaling for high-order moments. Varying the external forcing we are able to tune the system continuously from equilibrium, where the fluctuations are short-range correlated, to the case where the system is strongly driven in the infrared. In the latter case the nonequilibrium scaling of small-scale fluctuations are shown to be universal.}

\FullConference{31st International Symposium on Lattice Field Theory - LATTICE 2013\\
		July 29 - August 3, 2013\\
		Mainz, Germany}

\begin{document}

\section{Introduction}

Turbulence, the transport of conserved quantities in a strongly-correlated state far from equilibrium together with its universal scaling properties defines one of the most pressing problems in physics with relevance to systems so vastly different as, e.g., ultracold atomic gases at temperatures of a few nanokelvin or heavy ion collisions at ultra-relativistic energies. Its complexity can be overwhelming -- its study typically requires large computational efforts to straddle the huge range of scales that defines the problem. In such a situation, it is desirable to have a model at hand that allows for the important questions to be asked but removes all unnecessary details from the problem that would otherwise complicate its study. A much-employed example is the one-dimensional Burgers' equation without pressure \cite{Burgers:1973}. The pressureless Burgers' equation is free from the severe nonlocal interactions present in the incompressible Navier-Stokes equations usually taken to describe realistic hydrodynamic flow. Nevertheless, the presence of strong correlations over a wide range of scales induced by a power-law forcing makes this system sufficiently interesting from a physics perspective (see, e.g., Ref.~\cite{Bec:2007} for a review).

\section{Burgers' equation}

The random-force-driven Burgers' equation
\begin{equation}
\partial_{t} u + u \nabla u - \nu_{0} \nabla^{2} u = f(x,t)~,
\label{burgers}
\end{equation}
was originally conceived as a one-dimensional model for compressible hydrodynamic turbulence \cite{Burgers:1973} and provides a useful benchmark setting to test new analytical and numerical methods for real-world turbulence \cite{Frisch:2000,Bec:2007}. We consider the special case where the system is driven by a self-similar Gaussian forcing that is white in time. Its two-point correlation function in Fourier space is given by
\begin{equation}
\langle f(k ,t ) f(k' ,t') \rangle = 2 D_{0} |k|^{3-y} \delta(k+k') \delta(t-t') ~,
\label{eq:ForcingCorrelator}
\end{equation}
where the parameter $y$ determines the relative importance of fluctuations put into the system at different scales, and the dimensionful constant $D_{0}$ measures its strength. While large positive values of $y$ lead to a forcing that acts predominantly in the infrared (IR) in the opposite case, where $y$ is negative, the system is strongly driven in the ultraviolet (UV). Independent of the forcing mechanism, kinematic viscosity $\nu_{0}$ provides a dissipation scale $\eta$ and for $\nu_{0} \rightarrow 0$ the two characteristic scales $\eta$, and the finite system size $L$ separate. In that case, the stochastic forcing drives the system into a nonequilibrium steady state, where in the range \mbox{$\eta \ll k^{-1} \ll L$} the energy flux through wavenumber $k$ behaves as \mbox{$\Pi_{\varepsilon} (k) \sim k^{4 - y}$}. In contrast, for nonzero values of the viscosity a noise term acting in the UV can be used to model small frequency and long wavelength fluctuations in a fluid close to thermal equilibrium ($\Pi_{\varepsilon} = 0$) \cite{Forster:1977zz}. Thus, the parameter $y$ serves to control the type of scaling behavior -- depending on its value the character of excitations in the system will be very different. While the large-scale dominated forcing leads to the appearance of coherent shocks \cite{Chekhlov:1995aa} in the short-range correlated regime the system features no such structures and, in the long-time limit, is completely characterized by thermal noise.

The classical field-theoretic action for the random-force-driven Burgers' equation is obtained via the Martin-Siggia-Rose formalism \cite{Martin:1973zz,Janssen:1976,Phythian:1977}. Introducing the auxiliary response field $\tilde{u}$, we obtain the partition function
$Z = \int \left[d \tilde{u}\right]\! \left[d u\right] \, e^{-S}$
with the classical action
\begin{eqnarray}
S = \int_{[t_{0},\infty]}\! dt\, dx\, \left\{ \tilde{u} \left( \partial_{t} u + u \nabla u - \nu_{0} \nabla^2 u \right) - D_{0} \, \tilde{u} ( -\nabla^{2} )^{(3-y)/2} \tilde{u} \right\} ~,
\label{eq:ClassicalActionMSR}
\end{eqnarray}
where the quadratic noise term $\sim \tilde{u} |\nabla|^{3-y} \tilde{u}$ models the fluctuations that are put into the system by the stochastic forcing \eqref{eq:ForcingCorrelator}. Note, in the special case where $y = 1$ the one-dimensional Burgers' equation can be mapped to the Kardar-Parisi-Zhang equation \cite{Kardar:1986} which admits a generalized fluctuation dissipation relation (FDR). For generic values of $y > 0$, however, fluctuations are equally important on all scales leading to strong correlations in the system and a FDR is generally absent. It is the absence of a FDR and the presence of Galilean symmetry that essentially determines the phenomenology of the system and leads to the complex scaling behavior \cite{Eyink:1994zz,Polyakov:1995}. 

\section{Scaling behavior}

The classical action \eqref{eq:ClassicalActionMSR} depends on a single dimensionless coupling constant $g_{0}^{2} = ( D_{0} / \nu_{0}^{3} ) \Lambda^{-y}$ defined at the ultraviolet scale $\Lambda$, which is given in terms of the dimensionful force amplitude $D_{0}$ and the kinematic viscosity $\nu_{0}$. It naturally appears in a perturbative treatment of the problem (see, e.g., \cite{Forster:1977zz}), where one typically considers the following rescaling of the fields $t\rightarrow \nu_{0} t$, $u\rightarrow ( \nu_{0} / D_{0} )^{1/2} u$, and $\tilde{u}\rightarrow ( D_{0} / \nu_{0} )^{1/2} \tilde{u}$, to endow the nonlinear interaction term $\sim \tilde{u} ( u \nabla u )$ in \eqref{eq:ClassicalActionMSR} with the coupling $g_{0}$. Apart from the coupling constant $g_{0}^{2}$, the ratio between the lattice scale $\Lambda$ and the infrared cutoff $1/L$ defined by the inverse lattice size provides for a second dimensionless quantity that we may control. Eventually, we will be interested in the scaling behavior of correlation functions in the range between $1/L \ll k \ll \Lambda$, where both limits $\Lambda\rightarrow \infty$ and $L\rightarrow \infty$ are taken at the end. Depending on the values of the renormalized coupling in the limit where both cutoffs are removed one might expect different fixed point solutions that lead to a universal scaling behavior.

If the coupling $g_{0}^{2}$ is nonzero, the interplay of the nonlinearity with the Gaussian driving term leads to the most interesting dynamics. For scaling behavior of stationary nonequilibrium states one typically considers Galilei-invariant quantities, e.g., moments of field differences $\delta_{r}u = u(x+r) - u(x)$, where the value of the separation $r$ is small compared to the infrared scale $L$. Thus, we are concerned with the UV scaling properties in contrast to the IR scaling for excitations of a fluid near equilibrium \cite{Forster:1977zz}. For stochastic driven system \eqref{burgers} with a power-law spectrum, we obtain in the case of a homogeneous and isotropic flow
\begin{equation}
\langle \delta_{r} u \rangle \sim D_{0}^{1/3} r^{-1 + y/3} ~,
\label{eq:StructureFunctionScalingForcingDominated}
\end{equation}
which follows simply from dimensional analysis and leads to the scaling spectrum $\langle (\delta_{r}u)^{n} \rangle \sim D_{0}^{n/3} r^{\zeta_{n}^{(0)}}$, with $\zeta_{n}^{(0)} = n (-1 + y/3)$. It is important to emphasize, that this result implies that the viscous scale is removed $\nu_{0}\rightarrow 0$ to leave only the dimensionful parameter $D_{0}$. The associated spectral energy density for this particular scaling solution is defined via the Fourier transform of the second moment $\langle (\delta_{r}u)^{2}\rangle$. We obtain
\begin{equation}
E(k) \sim D_{0}^{2/3} k^{1 - 2 y/3} ~,
\label{eq:EnergySpectrumForcingDominated}
\end{equation}
which depends on the continuous parameter $y$. Note, that eq.~\eqref{eq:EnergySpectrumForcingDominated} describes a Kolmogorov energy spectrum $E \sim D_{0}^{2/3} k^{-5/3}$ when $y = 4$. 

These scaling scenarios, eqs.\ \eqref{eq:StructureFunctionScalingForcingDominated} and \eqref{eq:EnergySpectrumForcingDominated}, relate to Gaussian fixed points for the dynamics. However, both experiments and numerical simulations indicate strong intermittency effects for high order moments $\langle (\delta_{r} u)^{n}\rangle$, where $n \gg 1$. Thus, we are led to ask if the system allows for additional non-Gaussian scaling solutions characterized by sizable anomalous dimensions $\eta_{n} = \zeta_{n} - \zeta_{n}^{(0)}$ for the relevant scaling operators. To determine these scaling corrections from first principles is a tremendous task. One might hope that lattice Monte Carlo methods may provide a possibility to calculate the scaling spectrum unambiguously.

\section{Lattice discretization and global regularity of solutions}

The choice of discretization in a lattice Monte Carlo approach \cite{Duben:2008,Mesterhazy:2011kr} is a subtle issue for real-time dynamics. While an appropriate discretization in time is important for the cancellation of the functional determinant that in principle appears in the derivation of the partition function \cite{Honkonen:2011}, it also controls the character of physical solutions to the dynamics. In fact, this is well-known from the direct numerical solution of first order partial differential equations, where certain discretization schemes simply do not yields globally regular solutions. The most prominent example for such a behavior is the forward-time centered-space discretization (FTCS) scheme for the linear advective equation (see, e.g., \cite{Press:1992zz}). On the other hand, other discretizations may provide a dynamics that is conditionally stable, depending on the choice of the parameters in the problem. These observations are typically based on a linear stability analysis of the equations of motion and generally cannot be applied to nonlinear systems. 

Nevertheless, we find that a similar constraint applies for our choice of backward-time/pre-point discretization \cite{Mesterhazy:2011kr}. The dynamics is only conditionally stable which relates to the value of the lattice viscosity. If the lattice viscosity is chosen to be larger than $\hat{\nu}_{0} \simeq 1/2$ the dynamics will always feature instabilities. In fact, this particular bound is well understood from a similar discretization of the diffusion equation \cite{Press:1992zz}. This immediately poses the question if these problems may be overcome by using implicit time-differencing schemes as one usually applies for direct numerical solvers of partial differential equations (see, e.g., \cite{Ames:1967}). In what sense such discretizations are optimal and may lead to \emph{unconditionally} stable dynamics on the lattice is left for future work.

\section{Lattice Monte Carlo algorithms}

In this work two different types of algorithms were employed, an improved local overrelaxation algorithm \cite{Adler:1981sn,Adler:1987ce} and a variant of the Hybrid Monte Carlo (HMC) algorithm \cite{Duane:1987de}. While both algorithms have been discussed at length in the literature in the context of equilibrium systems, here, we discuss necessary adaptions and their application for simulations of classical-statistical dynamics in the presence of a stochastic driving term. We investigate specific improvements to control the performance of the HMC at the example of the one-dimensional Burgers' equation and comment on the relation to the improved overrelaxation algorithm. Results obtained with the overrelaxation algorithm, including the scaling spectrum $\zeta_{n}$ for the moments of field differences, were reported in Ref.~\cite{Mesterhazy:2011kr}.

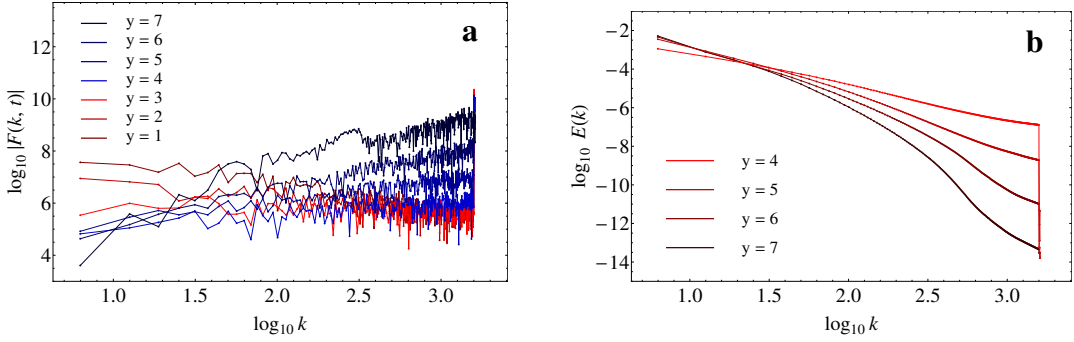
\begin{figure}
\centering
\includegraphics[width=0.44\textwidth]{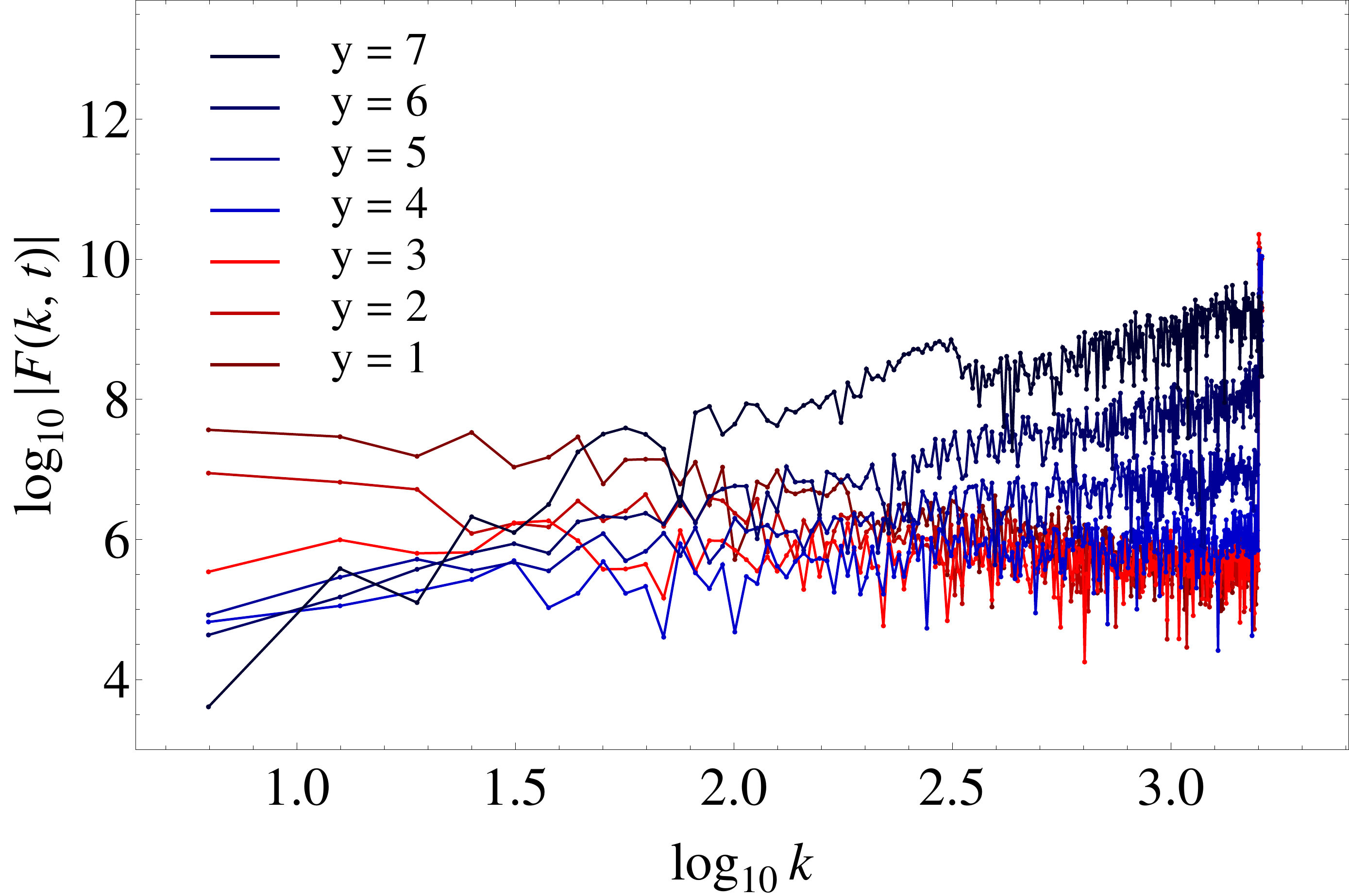}\qquad
\includegraphics[width=0.44\textwidth]{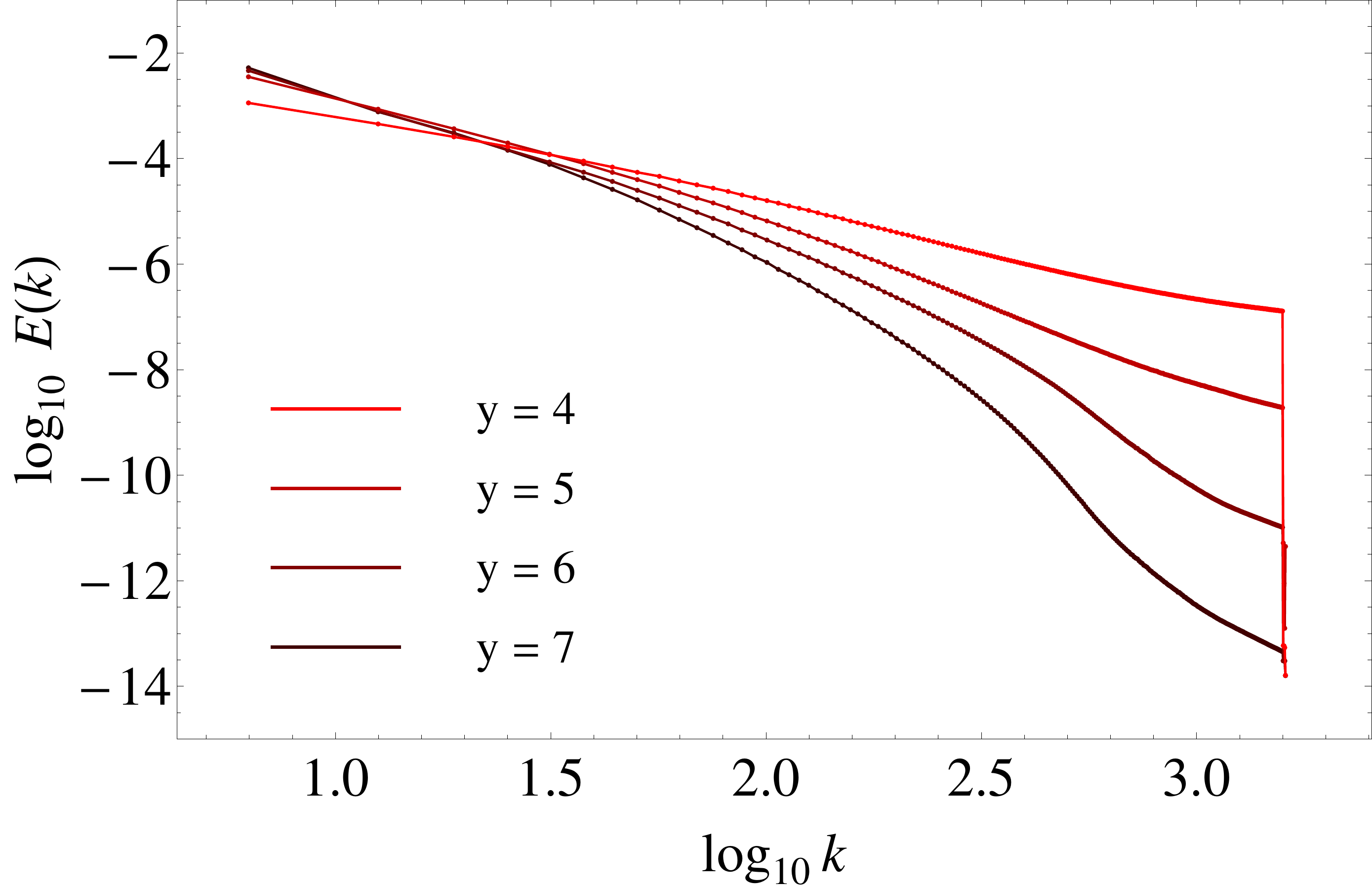}
\begin{picture}(300,-400)
\put(122,126){{\large \textbf{a}}}
\put(333,122){{\large \textbf{b}}}
\end{picture}
\caption{\label{fig:FourierSpectrumHMCForces}(\textbf{a}) MD force spectrum $| F(k,t) |$ at fixed physical time $t$ in a single $1024\times512$ (time$\times$space) configuration for different values of $y$. The data points are taken from a simulation where the Fourier acceleration is set iteratively after each Markov step. (\textbf{b}) Spectral energy density $E(k)$ evaluated on an ensemble of statistically independent configurations for different values of the exponent $y$.}
\end{figure}

A useful quantity that one may employ to monitor the performance of the HMC algorithm is the contribution $F = - \partial S / \partial u$ to the Molecular Dynamics (MD) evolution equations $\partial u/\partial \tau = \partial H_{\textrm{eff}} / \partial \pi$ and $\partial \pi/\partial \tau = - \partial H_{\textrm{eff}} / \partial u \equiv F$, generated by the effective Hamiltonian $H_{\textrm{eff}} = \frac{1}{2} ||\pi||^{2} + S(u)$, where $\tau$ is the MD evolution parameter and $\pi$ are the conjugate momenta. Typically, the system will strongly emphasize certain modes, while others are slowed down in comparison. This is illustrated in Fig.~\ref{fig:FourierSpectrumHMCForces}a where we show a typical sample of the MD forces $| F(k,t) |$ in the Fourier representation for a fixed value of the physical time $t$ measured on a single configuration. Clearly, the power-law forcing \eqref{eq:ForcingCorrelator} induces strong variations in the MD force spectrum, where $\overline{| F(k,t) |} \sim k^{(y-3)/2}$ on average, for $y = 1, \ldots, 7$. What is even more striking are the strong fluctuations induced by the real-time dynamics which cover a range of roughly two orders of magnitude. This makes the numerical solution quite demanding as the integrator has to tackle these different scales and avoid possible instabilities \cite{Edwards:1996vs} triggered by a too large value of the stepsize $\Delta \tau$ of the MD solver. In fact, a naive application of the HMC will not work unless the stepsizes are chosen extremely small, which might stabilize the integrator but effectively freezes the HMC dynamics.

Fourier acceleration \cite{Davies:1987vs} addresses this problem and suggests an alternative MD update which is based on the modified effective Hamiltonian $H_{\textrm{eff}}^{\textrm{FACC}} = \frac{1}{2} \sum_{x,x'; t} \pi(x,t) \Omega(x,x';t) \pi(x',t) + S( u )$. This choice yields an improved sampling of the conjugate momenta, and adapts the MD stepsizes for a given mode $\Delta \tau\rightarrow [ \Omega(k,t) ]^{1/2} \Delta \tau$. The kernel $\Omega$ is chosen such that it compensates for the strong fluctuations, i.e., we set $\Omega(k,t) = C_{\Omega} |F(k,t)|^{-1/2}$ after each Markov step iteratively with a proportionality factor $C_{\Omega}$ that requires tuning. We should point out, that an effectively field-dependent $\Omega(k) = \Omega(k;u)$ adapted at each step in the Markov process might alter the convergence properties of this HMC algorithm. In practice, one must check if the right fixed point distribution is sampled.

With these observations we have chosen to implement an adaption of the HMC which is local in time but global in space with a field-dependent sampling of the conjugate momenta. That is, for each physical time in a configuration a Metropolis step is carried out separately. The \emph{quasi-locality} of the algorithm requires an even-odd type update for the fields in the physical time direction. The presented adaptions have proven to be sufficient to yield a stable HMC for stochastic driven systems. It is worth pointing out that there are no comparable difficulties relating to the stability of the overrelaxation algorithm \cite{Duben:2008,Mesterhazy:2011kr}. In fact, it has been shown that the overrelaxation algorithm is competitive with a stochastic optimal Fourier accelerated Langevin-type algorithm \cite{Adler:1987ce}.

In Fig.~\ref{fig:FourierSpectrumHMCForces}b we give an overview on the data obtained so far with the quasi-local HMC. The shown energy spectra indicate a transition in the UV scaling behavior. For $y \gtrsim 6$ the scaling properties of the small-scale fluctuations deviate from the power-law behavior dictated by the Gaussian forcing. The current data suggest a crossover to a large-scale dominated forcing regime where the UV scaling behavior is expected to be universal. However, a careful analysis of the scaling contributions to $E(k)$ is necessary to make firm statements. This is currently in progress and results will be presented in a forthcoming publication \cite{Mesterhazy:2013}.

\section{Conclusions}

We have shown that lattice Monte Carlo simulations of driven nonequilibrium systems using a HMC algorithm are possible but require a careful set up. The presently employed quasi-local HMC with optimal tuning outperforms the overrelaxation algorithm \cite{Duben:2008,Mesterhazy:2011kr} considerably. This is not surprising as the strong coupling of a large number of degrees of freedom as present in hydrodynamic turbulence is hard to tackle with a fully local algorithm. In addition, the HMC allows for the simple inclusion of additional constraints in the microscopic action that might enable an improved importance sampling, that is, an efficient sampling of field configurations that contribute significantly to a specific class of observables. This is especially interesting since it was suggested that instanton configurations (rare events on the attractor) might play an important role to explain the asymptotic behavior of probability distribution functions and for the scaling behavior of high-order moments \cite{Balkovsky:1997zz}. Possible techniques in this direction are currently under investigation.

\acknowledgments

D.\,M. acknowledges discussions with S.~Mathey. We thank the bwGRiD, member of the German D-Grid initiative, CINECA, and the Scientific Computing Center at DESY Zeuthen for the availability of high performance computing resources and support.

\bibliography{references}

\begin{thebibliography}{10}

\bibitem{Adler:1981sn}
S.~L. Adler,
\newblock {\em {An overrelaxation method for the Monte Carlo evaluation of the
  partition function for multiquadratic actions}},
\newblock {\em Phys.\ Rev.} {\bf D23} (1981) 2901.

\bibitem{Adler:1987ce}
S.~L. Adler,
\newblock {\em {Overrelaxation algorithms for lattice field theories}},
\newblock {\em Phys.\ Rev.} {\bf D37} (1988) 458.

\bibitem{Ames:1967}
W.~F. Ames,
\newblock {\em {Nonlinear partial differential equations}},
\newblock Academic Press (1992).

\bibitem{Balkovsky:1997zz}
E.~Balkovsky, G.~Falkovich, I.~Kolokolov, and V.~Lebedev,
\newblock {\em {Intermittency of Burgers' Turbulence}},
\newblock {\em Phys.\ Rev.\ Lett.} {\bf 78} (1997) 1452
  [\href{http://xxx.lanl.gov/abs/chao-dyn/9609005}{{\ttfamily
  chao-dyn/9609005}}].

\bibitem{Bec:2007}
J.~Bec and K.~Khanin,
\newblock {\em {Burgers turbulence}},
\newblock {\em Phys.\ Rep.} {\bf 447} (2007) 1
  [\href{http://xxx.lanl.gov/abs/0704.1611}{{\ttfamily 0704.1611}}].

\bibitem{Burgers:1973}
J.~M. Burgers,
\newblock {\em {The nonlinear diffusion equation: asymptotic solutions and
  statistical problems}},
\newblock D.\ Reidel Pub.\ Co. (1973).

\bibitem{Chekhlov:1995aa}
A.~Chekhlov and V.~Yakhot,
\newblock {\em {Kolmogorov turbulence in a random-force-driven Burgers
  equation}},
\newblock {\em Phys.\ Rev.} {\bf E51} (1995) R2739
  [\href{http://xxx.lanl.gov/abs/adap-org/9506002}{{\ttfamily
  adap-org/9506002}}].

\bibitem{Davies:1987vs}
C.~Davies, G.~Batrouni, G.~Katz, A.~S. Kronfeld, G.~Lepage, et~al.,
\newblock {\em {Fourier acceleration in lattice gauge theories. 1. Landau gauge
  fixing}},
\newblock {\em Phys.\ Rev.} {\bf D37} (1988) 1581.

\bibitem{Duane:1987de}
S.~Duane, A.~Kennedy, B.~Pendleton, and D.~Roweth,
\newblock {\em {Hybrid Monte Carlo}},
\newblock {\em Phys.\ Lett.} {\bf B195} (1987) 216.

\bibitem{Duben:2008}
P.~D{\"u}ben, D.~Homeier, K.~Jansen, D.~Mesterh{\'a}zy, G.~M{\"u}nster, and
  C.~Urbach,
\newblock {\em {Monte Carlo simulations of the randomly forced Burgers
  equation}},
\newblock {\em Europhys.\ Lett.} {\bf 84} (2008) 40002
  [\href{http://xxx.lanl.gov/abs/0809.4959}{{\ttfamily 0809.4959}}].

\bibitem{Edwards:1996vs}
R.~Edwards, I.~Horvath, and A.~Kennedy,
\newblock {\em {Instabilities and nonreversibility of molecular dynamics
  trajectories}},
\newblock {\em Nucl.\ Phys.} {\bf B484} (1997) 375
  [\href{http://xxx.lanl.gov/abs/hep-lat/9606004}{{\ttfamily
  hep-lat/9606004}}].

\bibitem{Eyink:1994zz}
G.~L. Eyink,
\newblock {\em {The renormalization group method in statistical
  hydrodynamics}},
\newblock {\em Phys.\ Fluids} {\bf 6} (1994) 3063.

\bibitem{Forster:1977zz}
D.~Forster, D.~R. Nelson, and M.~J. Stephen,
\newblock {\em {Large-distance and long-time properties of a randomly stirred
  fluid}},
\newblock {\em Phys.\ Rev.} {\bf A16} (1977) 732.

\bibitem{Frisch:2000}
U.~Frisch and J.~Bec,
\newblock {\em {Burgulence}}
\newblock  [\href{http://xxx.lanl.gov/abs/nlin.CD/0012033}{{\ttfamily
  nlin.CD/0012033}}].

\bibitem{Honkonen:2011}
J.~Honkonen,
\newblock {\em {Ito and Stratonovich calculuses in stochastic field theory}}
\newblock  [\href{http://xxx.lanl.gov/abs/1102.1581}{{\ttfamily 1102.1581}}].

\bibitem{Janssen:1976}
H.-K. Janssen,
\newblock {\em {On a Lagrangean for classical field dynamics and
  renormalization group calculations of dynamical critical properties}},
\newblock {\em Z.\ Phys.} {\bf B23} (1976) 377.

\bibitem{Kardar:1986}
M.~Kardar, G.~Parisi, and Y.-C. Zhang,
\newblock {\em {Dynamic Scaling of Growing Interfaces}},
\newblock {\em Phys.\ Rev.\ Lett.} {\bf 56} (1986) 889.

\bibitem{Martin:1973zz}
P.~C. Martin, E.~D. Siggia, and H.~A. Rose,
\newblock {\em {Statistical Dynamics of Classical Systems}},
\newblock {\em Phys.\ Rev.} {\bf A8} (1973) 423.

\bibitem{Mesterhazy:2013}
D.~Mesterh{\'a}zy, L.~Biferale, K.~Jansen, and R.~Tripiccione,
\newblock {\em {To be published}}.

\bibitem{Mesterhazy:2011kr}
D.~Mesterh{\'a}zy and K.~Jansen,
\newblock {\em {Anomalous scaling in the random-force-driven Burgers' equation:
  A Monte Carlo study}},
\newblock {\em New J.\ Phys.} {\bf 13} (2011) 103028
  [\href{http://xxx.lanl.gov/abs/1104.1435}{{\ttfamily 1104.1435}}].

\bibitem{Phythian:1977}
R.~Phythian,
\newblock {\em {The functional formalism of classical statistical dynamics}},
\newblock {\em J.\ Phys.} {\bf A10} (1977) 777.

\bibitem{Polyakov:1995}
A.~M. Polyakov,
\newblock {\em {Turbulence without pressure}},
\newblock {\em Phys.\ Rev.} {\bf E52} (1995) 6183
  [\href{http://xxx.lanl.gov/abs/hep-th/9506189}{{\ttfamily hep-th/9506189}}].

\bibitem{Press:1992zz}
W.~H. Press, S.~A. Teukolsky, W.~T. Vetterling, and B.~P. Flannery,
\newblock {\em {Numerical Recipes: The Art of Scientific Computing}},
\newblock Cambridge University Press (2007).

\end{thebibliography}

\end{document}